\documentclass[]{aa}
\usepackage[acronym,toc,nonumberlist]{glossaries}

\newacronym{CNES}{CNES}{Centre National d'Etudes Spatiales}
\newacronym{DSN}{DSN}{Deep Space Network}
\newacronym{ESA}{ESA}{European Space Agency}
\newacronym{GINS}{GINS}{G\'eod\'esie par Int\'egrations Num\'eriques Simultan\'ees}
\newacronym{INPOP}{INPOP}{Int\'egrateur Num\'erique Plan\'etaire de l'Observatoire de Paris}
\newacronym{LOS}{LOS}{line of sight}
\newacronym{NAIF}{NAIF}{Navigation and Ancillary Information Facility}
\newacronym{NASA}{NASA}{National Aeronautics and Space Administration}
\newacronym{PDS}{PDS}{Planetary Data System}
\newacronym{SEP}{SEP}{Sun Earth Probe}
\newacronym{VEX}{VEX}{Venus Express}
\newacronym{MGR}{MESSENGER}{MErcury Surface, Space ENvironment, GEochemistry, and Ranging}
\newacronym{JPL}{JPL}{Jet Propulsion Laboratory} 
\newacronym{RMDCT}{RMDCT}{Radio Metric Data Conditioning Team}
\newacronym{ODF}{ODF}{Orbit Data File}
\newacronym{EOP}{EOP}{Earth Orientation Parameters}
\newacronym{IAU}{IAU}{International Astronomical Union}
\newacronym{HEF}{HEF}{mount high efficiency}
\newacronym{BVE}{BVE}{block 5 exciter}
\newacronym{COI}{COI}{center of integration}
\newacronym{PPN}{PPN}{Parametrized Post Newtonian}
\newacronym{ROB}{ROB}{Royal Observatory of Belgium}
\newacronym{rms}{rms}{root mean square}
\newacronym{GR}{GR}{general relativity}

\makeglossaries

\usepackage{footnote}
\usepackage{graphicx}
\usepackage{natbib}
\usepackage[utf8]{inputenc}
\usepackage{graphics}
\usepackage[subnum]{cases}
\usepackage{threeparttable}
\usepackage[titletoc,title]{appendix}
\usepackage[dotinlabels]{titletoc}
\usepackage{makecell}
\usepackage[colorlinks=true,linkcolor=red,citecolor=blue]{hyperref}
\usepackage{lscape}
\usepackage[figuresright]{rotating}
\usepackage{multirow,booktabs}
\usepackage{ulem}
\usepackage{txfonts}
\usepackage{colortbl}                          
\usepackage{stfloats}
\usepackage{caption}
\usepackage{subcaption}
\usepackage{enumerate}
\usepackage[b]{esvect }
\usepackage{textcomp}

\def\inst#1{\unskip$^{#1}$}

 \def\apjl{ApJ}%
 \def\apjs{ApJS}%
 \def\ao{Appl.~Opt.}%
 \def\aap{A\&A}%
\begin{document}
\title{Tests of General relativity with planetary orbits and Monte Carlo simulations}

   \author{A. Fienga\inst{1,2},  J. Laskar\inst{2}, P. Exertier\inst{1}, H. Manche\inst{2}, M. Gastineau\inst{2}}
   \institute{G\'eoazur, CNRS/UMR7329, Observatoire de la C\^ote d'Azur, Valbonne, France
	  \and
	    IMCCE, CNRS/UMR8028, Observatoire de Paris, Paris, France
            }

\date{\today}

 \titlerunning{Test of relativity with INPOP and Monte Carlo simulations}
 \authorrunning{Fienga et al.}

\abstract{Based on the new developped planetary ephemerides INPOP13c, determinations of acceptable intervals of General Relativity violation in considering simultaneously the PPN parameters $\beta$, PPN $\gamma$, the flattening of the sun $J_{2}^\odot$ and time variation of the gravitational mass of the sun $\mu$ are obtained in using  Monte Carlo simulation coupled with basic genetic algorithm. Possible time variations of the gravitational constant G are also deduced. Discussions are lead about the better choice of indicators for the goodness-of-fit for each run and limits consistent with general relativity are obtained simultaneously.

   }
   \maketitle

\section{Introduction}


Since the first publications of the theory of General Relativity (\cite{2007ggr..conf..543E}), the dynamics of the solar system has demonstrated its efficiency for testing gravity theories. Examples can be given such as the measurements  of the advance of perihelia (\cite{1859AnPar...5....1L}) of Mercury's orbit, firstly observed and then explained by General Relativity (GR), or of the asteroid Icarus's orbit (\cite{1965AJ.....70R.675F}), or the measurement of the deflexion of light by the Sun during the cruise phase of Cassini probe (\cite{2003Natur.425..374B}).
Direct implementations of alternative metrics in the equations of motion of the planets of the solar system and of spacecraft orbiting planets were also introduced (\cite{2008JGeod..82..133M}, \cite{2011Icar..211..401K}, \cite{2012CQGra..29w5027H}) stressing the interest of using the solar system as a laboratory for testing alternative modelings of gravity. 
Tests based on the analysis of MESSENGER spacecraft tracking data used in the construction of INPOP ephemerides or the Cassini tracking data for the JPL DE ephemerides  have also demonstrated the impact of using planetary ephemerides for eliminating possible values of parameters characterizing alternative theories of gravity (\cite{2010IAUS..261..159F}, \cite{2014A&A...561A.115V}, \cite{2014PhRvD..89j2002H}).

Planetary ephemerides can thus be used as a very efficient tool for several scientific applications. 
For example, with the INPOP planetary ephemerides developed since 2003, determinations of the flattening of the Sun $J_{2}^\odot$, of the PPN parameters $\beta$ and $\gamma$, and estimates of the perihelion advances of the 8 planets of the solar system have been obtained with considerable precision. No new advance and no $\beta$ and $\gamma$ deviations from unity could be exhibited, confirming the validity of general relativity to the level of 10$^{-5}$ (\cite{Fienga2011},\cite{2014A&A...561A.115V}). INPOP constraints on supplementary advances of nodes and perihelia of the planets give stringent constraints on cosmological models of gravity such as MOND (\cite{2011MNRAS.412.2530B}). The INPOP limits of violation for PPN $\beta$ and $\gamma$ are important constraints for certain type of tensor-scalar theories where cosmological evolution exhibits an attractor mechanism towards GRT (\cite{1993PhRvD..48.3436D}) or in string-inspired scalar-tensor theories where the scalar field can decouple from matter (\cite{1994GReGr..26.1171D}).

Finally, some  phenomenological tests are related to the constancy of the Newtonian gravitational constant G in time. Variations of G are produced e.g. by alternative theories of gravity like tensor-scalar theory (see e.g. \cite{1990PhRvL..64..123D} and \cite{2003AnHP....4..347U}) or by some models of Dark Energy (\cite{2009PhRvD..79j4026S}; \cite{2007PhRvD..75l3007F}). The $\dot{G}/G$ ratio is now constrained at the level of 10$^{-13}$ with LLR analysis (\cite{2004PhRvL..93z1101W}) but no tests combining simultaneously modifications of PPN formalism, variation of the sun flattening and of the Newtonian gravitational constant were done. 
Some estimations of PPN violations and of $\dot{G}$ were obtained by \cite{2013MNRAS.432.3431P}. 

The purpose of this work is to make the first simultaneous determinations of possible variations from unity of PPN parameters $\beta$, $\gamma$ and of non-zero value for $\dot{G}$ but also in including the determination of the flattening of the Sun. We consider the equation of the advance in perihelia $\Delta\dot\varpi$ for any planet of the solar system in the PPN formalism as a good tool for monitoring the question of the determination of these parameters. We have: 
\begin{equation}
\Delta\dot\varpi = \frac{2\pi(2\gamma-\beta+2) GM_{\odot}(t)}{a(1-e^2)c^2}+\frac{3\pi J_{2}^{\odot} R^2_{\odot}}{a^2(1-e^2)c^2} + \Delta\dot\varpi_{AST}
\label{perih}
\end{equation}
where $a$ and $e$ are the semi-major axis and the eccentricity of the planet orbit respectively,  $R^2_{\odot}$ is the Sun radius respectively and $\Delta\dot\varpi_{AST}$ is the advance of the planet perihelia induced by the perturbations of the asteroids.
The PPN term is directly related to the Sun flatness J$_{2}$$^{\odot}$ but also to any variation of G through $GM_{\odot}(t)$. The term induced by the asteroid perturbations will be taken care by the iterative fit of the ephemerides to the observations and by the determination of the masses of the perturbing asteroids. The disentangling of the  PPN terms with Sun J$_{2}$$^{\odot}$ will be reinforced by the use of the eight orbits of the solar system planets and the introduction of eight different values for the semi-major axis $a$.
A non-unity PPN $\gamma$ is also introduced in the equation of the deflexion of light required for the computation of residuals as well as in the analysis of spacecraft tracking data. 
However as it was demonstrated in \cite{VermaPhD}, the orbital variations induced by the change in the values of the PPN $\gamma$ and $\beta$ parameters can be absorbed during the fit and the construction of the spacecraft orbit. No iteration is then required and only the impact of the modified deflexion of light on the deduced planetary light time has to be taken into account.

By considering simultaneously the construction of planetary ephemerides built with values of PPN parameters different from unity, with a gravitational constant G varying with time and a non-zero Sun flattening, we obtain the first consistent tests of this type based on planetary ephemerides. However, because of the important number of simulations required by this approach, we use optimization method and more specifically genetic algorithm.
Since their first implementations in astrophysics by \cite{1995ApJS..101..309C}, the genetic algorithms and the evolutionary computation have been intensively used in the dynamical studies of the extra-solar systems (\cite{2013ApJ...770...53N}, \cite{2012MNRAS.427..770M}, \cite{2012MNRAS.425..749H}, \cite{2012MNRAS.425..930G}) but also in other fields such as astrosismology, binary physics, planetology  (\cite{2013AJ....146..149K}, \cite{2014A&A...565A..93S}, \cite{2013Icar..226..212H}). 
For the extra-solar planetary systems, the goal is to determine which combinations of planets can generate a radial velocity signature very similar to the one observed. 
Due to the constraints imposed by the dynamics, the algorithms applied for this type of studies can be very complex (\cite{2012RAA....12.1044B}). 
In our case, only starting intervals for the varying parameters (PPN $\beta$, $\gamma$, $J_{2}^\odot$ and $\dot{\mu}/\mu$) are given and a simple two crossovers algorithm is used. The applied genetic algorithm is fully described in section \ref{GAintro} and several fitness functions are discussed in sections \ref{GAintro} and \ref{GAdisc}. Corresponding results are presented in sections \ref{GAresults1} and \ref{GAresults2}.

\section{Implementation of GR tests in INPOP13c}
\label{intro}
The computations performed in this work are based on the new version of the INPOP ephemerides, INPOP13c. The following section introduces INPOP13c and the necessary modifications of the equations of motion in order to introduce the time variations of the Newtonian gravitational constant.

\subsection{Description of INPOP13c}
\label{inpop13b}

By the use of the tracking data of the MESSENGER mission, INPOP13a becomes an important tool for testing General Relativity close to the Sun (\cite{2014A&A...561A.115V}).  
The MESSENGER mission was indeed the first mission dedicated to the study of Mercury. 
The spacecraft orbits the smallest and closest to the sun planet of the solar system since 2011.
In (\cite{2014A&A...561A.115V}) are described the methods and procedures used for the analysis of the MESSENGER Doppler and range data included in the construction of the Mercury improved ephemerides, INPOP13a as well as the determination of acceptable intervals of non-unity values for the PPN $\beta$ and $\gamma$.

INPOP13c is an upgraded version of INPOP13a, fitted to LLR observations, and including new observables of Mars and Venus deduced from MEX, Mars Odyssey and VEX tracking (\cite{M2012}, \cite{M2013} , \cite{Marty13}). Table \ref{omctab1} of appendix A resumes the data samples and the obtained residuals with INPOP13c and INPOP10e common to the two ephemerides when Table \ref{omcsupp} exhibits the residuals obtained for the data samples added since INPOP10e.
Thanks to this supplementary material, INPOP13c faces a better extrapolation capability of the Mars ephemerides  as well as better consistencies between DE and INPOP ephemerides. For more details, tables and plots can be found in \cite{INPOP13c}.

The INPOP13c adjustment of the gravitational mass of the sun was performed as recommended by the IAU resolution B2 as well as the sun oblateness (J$_{2}$$^{\odot}$), the ratio between the mass of the earth and the mass of the moon (EMRAT) and the mass of the Earth-Moon barycenter. Estimated values are presented on Table \ref{paramfita}.

\begin{table*}
\caption{Values of parameters obtained in the fit of INPOP13c, INPOP10e and DE430 (\cite{2014IPNPR.196C...1F}) to observations.}
\begin{center}
\begin{tabular}{l c c c }
\hline
&  INPOP13c & INPOP10e & DE430 \\
&    $\pm$ 1$\sigma$ & $\pm$ 1$\sigma$ & $\pm$ 1$\sigma$ \\
\hline
(EMRAT-81.3000)$\times$ 10$^{-4}$ &  (5.694 $\pm$ 0.010) & (5.700 $\pm$ 0.020) & ($\pm$) \\

J$_{2}$$^{\odot}$ $\times$ 10$^{-7}$ & (2.30 $\pm$ 0.25) & (1.80 $\pm$ 0.25)  & 1.80  \\
\hline
GM$_{\odot}$ - 132712440000 [km$^{3}.$ s$^{-2}$]&  (44.487 $\pm$ 0.17) & (50.16 $\pm$ 1.3) & 40.944 \\
AU - 1.49597870700 $\times$ 10$^{11}$ [m] & 0.0 & 0.0 & (-0.3738 $\pm$ 3 ) \\
\hline
 [M$_{\odot}$ / M$_{\textrm{EMB}}$] - 328900 & 0.55314 $\pm$ 0.00033  &  0.55223 $\pm$ 0.004  &  0.55915 $\pm$ NC \\
 \hline
 \end{tabular}
\end{center}
\label{paramfita}
\end{table*}

\subsection{Implementation of GR tests}
\subsubsection{PPN $\beta$, $\gamma$ and $J_{2}^\odot$}

Based on the procedures described in \cite{2010IAUS..261..159F}, \cite{Fienga2011} and \cite{2014A&A...561A.115V}, evaluations of acceptable values of PPN $\beta$ and $\gamma$ were obtained for previous versions of INPOP by comparisons of the postfit residuals obtained with planetary ephemerides numerically integrated and fitted with values of $\beta$ and $\gamma$ different from unity.
Such procedure was set up to answer to questions rose by alternative theories of gravitation proposing solar systems facing non-zero PPN parameters. As described in \cite{Fienga2011} and \cite{2014A&A...561A.115V}, a grid of sensitivity of PPN $\beta$ and $\gamma$ but also of supplementary advances in the perihelia and the nodes of the planets was set up and compared to observations. 
Such a grid is built by constructing ephemerides with non-unity PPN parameters, numerically integrated and fitted to the same data sample as the regular INPOP. We then consider the maximum differences between the postfit residuals obtained with the ephemerides built in a non relativistic frame and INPOP. 
The limit chosen for defining the acceptable intervals of PPN parameters leading to postfit residuals close to INPOP was largely discussed in \cite{2014A&A...561A.115V} leading to an optimized value of 25\%. 

\subsubsection{Variation of the gravitational constant}
\label{gdot}
For this work, we added in the INPOP dynamical model the possibility of constraining variations of the gravitational mass of the sun, $\mu$, considering a variation of the mass of the sun noted $\dot{M_{\odot}}$ (often interpreted in stellar physics as the Sun mass loss) and a variation of the gravitational constant $\dot{G}$. We have with $\mu = G \times M_{\odot}$:

\begin{equation}
 \frac{\dot{\mu}}{\mu} = \frac{\dot{G}}{G} + \frac{\dot{M_{\odot}}}{M_{\odot}} 
\end{equation}

At each step $t$ of the numerical integration  of the INPOP equations of motion, we then estimate :
\begin{eqnarray}
M_{\odot}(t) &=& M_{\odot}(J2000) + (t-J2000) \times \dot{M_{\odot}} \\
G(t) &=& G(J2000) + (t-J2000) \times \dot{G}\\
\mu(t)&=& G(t)  \times M_{\odot}(t) 
\label{mudot}
\end{eqnarray}

$\dot{\mu}/{\mu}$ plays also a role in the computation of the Shapiro delay of the observables (see \cite{Moyer2000}). 
In this case, the value of $\mu$ corresponding to the date of the observation is computed with Equation \ref{mudot} and re-introduced in the Shapiro delay equation (8-38) given in \cite{Moyer2000}.

Tests are then operated by considering different values of $\dot{\mu}/{\mu}$. We use for $M_{\odot}(J2000)$ the fitted mass of the sun as estimated for each planetary ephemerides (the value obtained for INPOP13c is given in Table \ref{paramfita}) and for $G(J2000)$ the Newtonian gravitation constant as defined by the IAU (\cite{2011CeMDA.110..293L}). 
We then deduce values of $\dot{G}/G$ in considering a fixed value for the Sun total mass loss (including radiation and solar winds):
 $$\frac{\dot{M_{\odot}}}{M_{\odot}} = (-0.55 \pm 0.15) \times 10^{-13} \, \textrm{yr}^{-1} $$
 extracted from solar physics measurements and variations of $\dot{M_{\odot}}/{M_{\odot}}$ during the 11-year solar cycle (\cite{2011ApJ...737...72P}).

\section{Monte Carlo simulations}

\subsection{Principle}
\label{3.1}
\begin{figure}
\includegraphics[width=8cm]{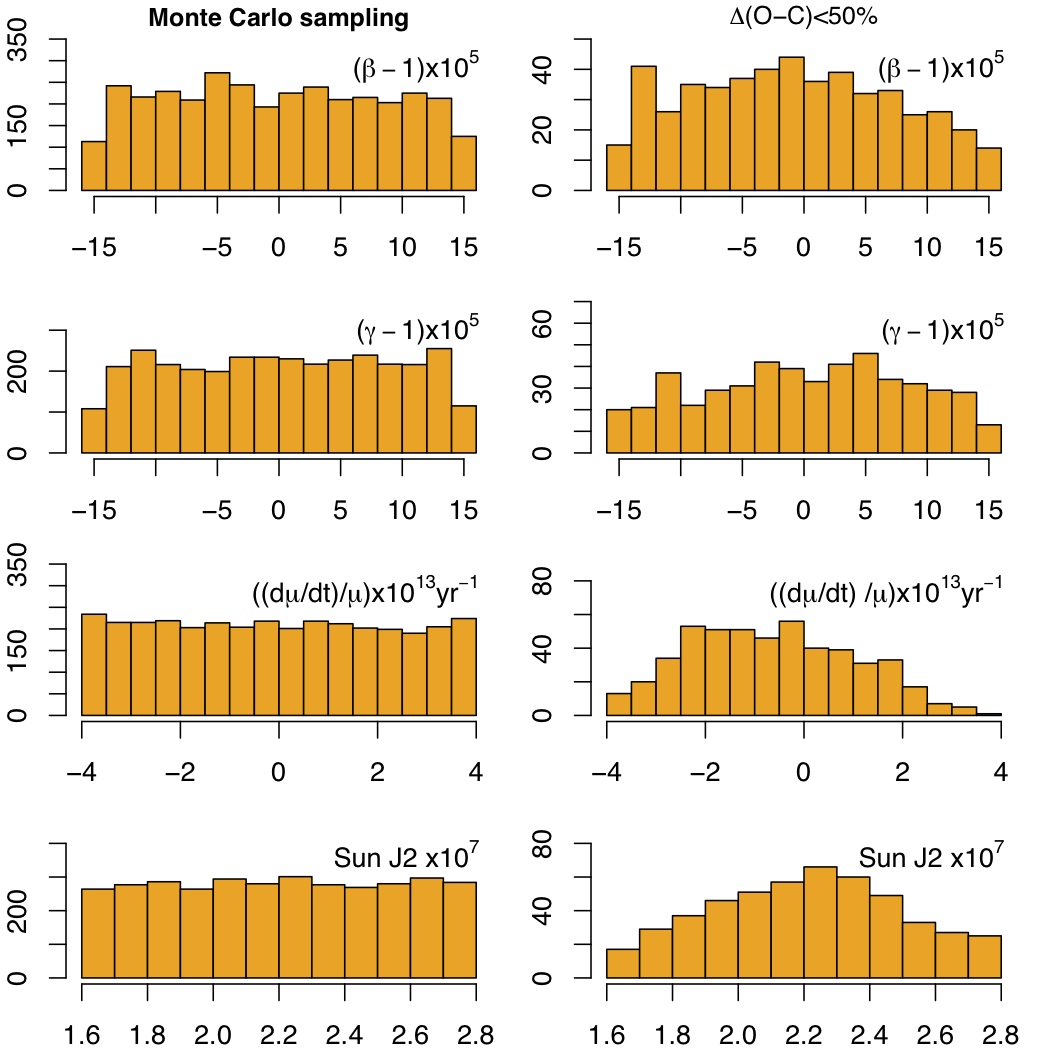}
\caption{On the left: histograms of the uniform random values for PPN $\beta$, PPN $\gamma$, $J_{2}^\odot$, and $\dot{\mu}/\mu$ parameters used for the construction of the planetary ephemerides for the Monte Carlo simulation. On the right: histograms of the distribution of the PPN $\beta$, PPN $\gamma$, $J_{2}^\odot$, and $\dot{\mu}/\mu$ parameters corresponding to the selected ephemerides using the criteria of 50\% differences of postfit residuals. See section \ref{3.2}}
\label{MC}
\end{figure}

Following the strategy described above, an uniform sampling of PPN $\beta$, $\gamma$, $J_{2}^\odot$ and $\dot{\mu}/\mu$ was built in order to be used for the construction of planetary ephemerides adjusted to observations. 
Postfit residuals were obtained after several iterations and compared to those obtained with the nominal planetary ephemerides INPOP13c built in the GR framework with $\beta=\gamma=1 \, , \, \dot{\mu}/\mu=0$ and $J_{2}^\odot=2.3$ as given in Table \ref{paramfita}.
The plots of the left-hand side of Figure \ref{MC} show the uniform distribution of the GR parameters obtained after 4000 runs. The right-hand side plots give the histograms of one possible selection of ($\beta$, $\gamma$, $J_{2}^\odot$, $\dot{\mu}/\mu$) (see section \ref{3.2}).
The intervals of the uniform distributions were chosen larger than the present limits found in the literature and gathered in Table \ref{tab1}. For the PPN parameters $\beta$ and $\gamma$, we selected an interval of $\pm 15 \times 10^{-5}$ enclosing VLBI and LLR determinations (\cite{2009IJMPD..18.1129W}, \cite{2009AA...499..331L}). For the Sun flattening, our interval from $1.6 \times 10^{-7}$ to $2.8 \times 10^{-7}$ also includes the values found in the literature mainly extracted from planetary ephemerides construction (\cite{2014A&A...561A.115V}, \cite{2009AA...507.1675F}).

\subsection{First Results}
\label{3.2}
After the constructions of the planetary ephemerides based on an uniform sampling of values for ($\beta$, $\gamma$, $J_{2}^\odot$, $\dot{\mu}/\mu$), postfit residuals were obtained and compared with those given by INPOP13c. 
Criteria are then chosen for selecting values of ($\beta$, $\gamma$, $J_{2}^\odot$, $\dot{\mu}/\mu$) leading to ephemerides close to INPOP13c. For these selected values, it is not possible to say in the light of the present accuracy of the observations if they induced noticeable violation of planetary dynamics. 
A possible approach for the selection of such ephemerides and ($\beta$, $\gamma$, $J_{2}^\odot$, $\dot{\mu}/\mu$) sampling is to consider the maximum differences noted $\Delta (O-C)_{max}$ between the postfit residuals of the non-GR ephemerides  and those of INPOP13c. This strategy was used previously in \cite{Fienga2011} and \cite{2014A&A...561A.115V} and lead to a threshold of 25\%  for $\Delta (O-C)_{max}$ as an appropriate limit for considering noticeable violations of planetary dynamics. 

Because of the increase of the number of free parameters in the present simulations compared to the previous publications where only PPN $\beta$ and $\gamma$ were randomly modified, the postfit difference threshold was rose up to 50\% in a first attempt to define violation limits. Only 15\% of the 4000 runs give postfit residual differences to INPOP13c below 50\% and only 1.4\% runs give differences below 25\% .
On the right-hand side on Figure \ref{MC}, the distributions of values for $\beta$, $\gamma$, $J_{2}^\odot$ and $\dot{\mu}/\mu$ parameters corresponding to 15 \% of the selected ephemerides are plotted.
As it appears, except perhaps for $J_{2}^\odot$, the selected values do not show a gaussian distribution even after postfit residual selections.
Because on these results, an optimization of the Monte Carlo simulations based on a genetic algorithm was set up.

\section{Optimized MC simulations with Genetic Algorithm}
\subsection{Principle}
\label{GAintro}
\begin{figure}
\includegraphics[width=8cm]{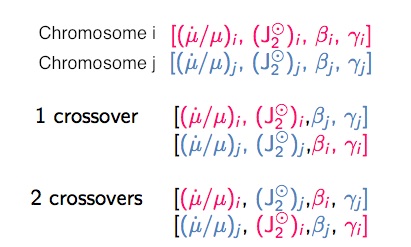}
\caption{Scheme of the genetic algorithm procedure. }
\label{algogen}
\end{figure}

We consider a genetic algorithm with two crossovers\footnote{Genetic operator used for varying the parameters from one generation to another. In our case, a simple swap is used.\cite{Magnin98}} and a probability of mutation of 10 \% as illustrated on Figure \ref{algogen}. 
The considered chromosome\footnote{set of parameters defining a proposed solution to the problem and randomly modified by the algorithm. \cite{Magnin98}} is here one set of four values of ($\beta$, $\gamma$, $J_{2}^\odot$, $\dot{\mu}/\mu$) and the individual is an INPOP planetary ephemerides built in using the values of ($\beta$, $\gamma$, $J_{2}^\odot$, $\dot{\mu}/\mu$), numerically integrated and fitted to observations in an iterative process. 
The fitness of each individual is the difference of its postfit residuals to INPOP13c noted $\Delta (O-C)$.

A total of 35800 runs spread over 30 generations were computed on the Paris Science et Lettres mesocentre with NEC 1472 kernels on 92 nodes. Two nodes were allocated for this work allowing 12 runs over 1 hour, one run representing the construction of about four numerically integrated and iteratively fitted planetary ephemerides. 
The 4000 runs of the Monte Carlo simulation presented in section \ref{3.1} were used to initialize the algorithm as the generation 0.

\begin{figure}
\includegraphics[width=8cm]{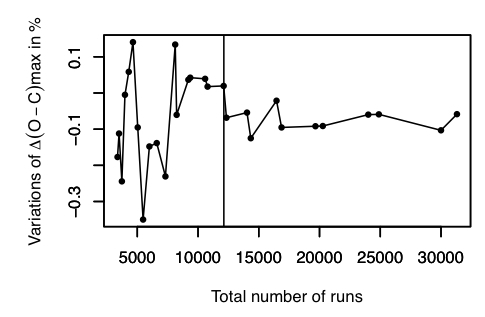}
\caption{Convergence of the Genetic Algorithm. The x-axis is the total number of runs accumulated over the generations while on the y-axis are plotted the differences for the average maximum differences $\Delta (O-C)_{max}$ in post residuals obtained for each generation. The dashed line indicates the 18th generation reaching a total number of 12125 runs.}
\label{conv}
\end{figure}

We stop the generational process until the average change in the maximum differences of postfit residuals $\Delta (O-C)_{max}$ is stabilized.
As one can see on Figure \ref{conv}, these differences stabilized at about 12 000 runs corresponding to the 18th generation. 
A clear change of regime is then obtained with a dispersion of $0.14 \% $ before the generation 18 and of $0.03 \%$ after. 

\subsection{Results}
\label{GAresults1}

\begin{figure}
\includegraphics[width=8cm]{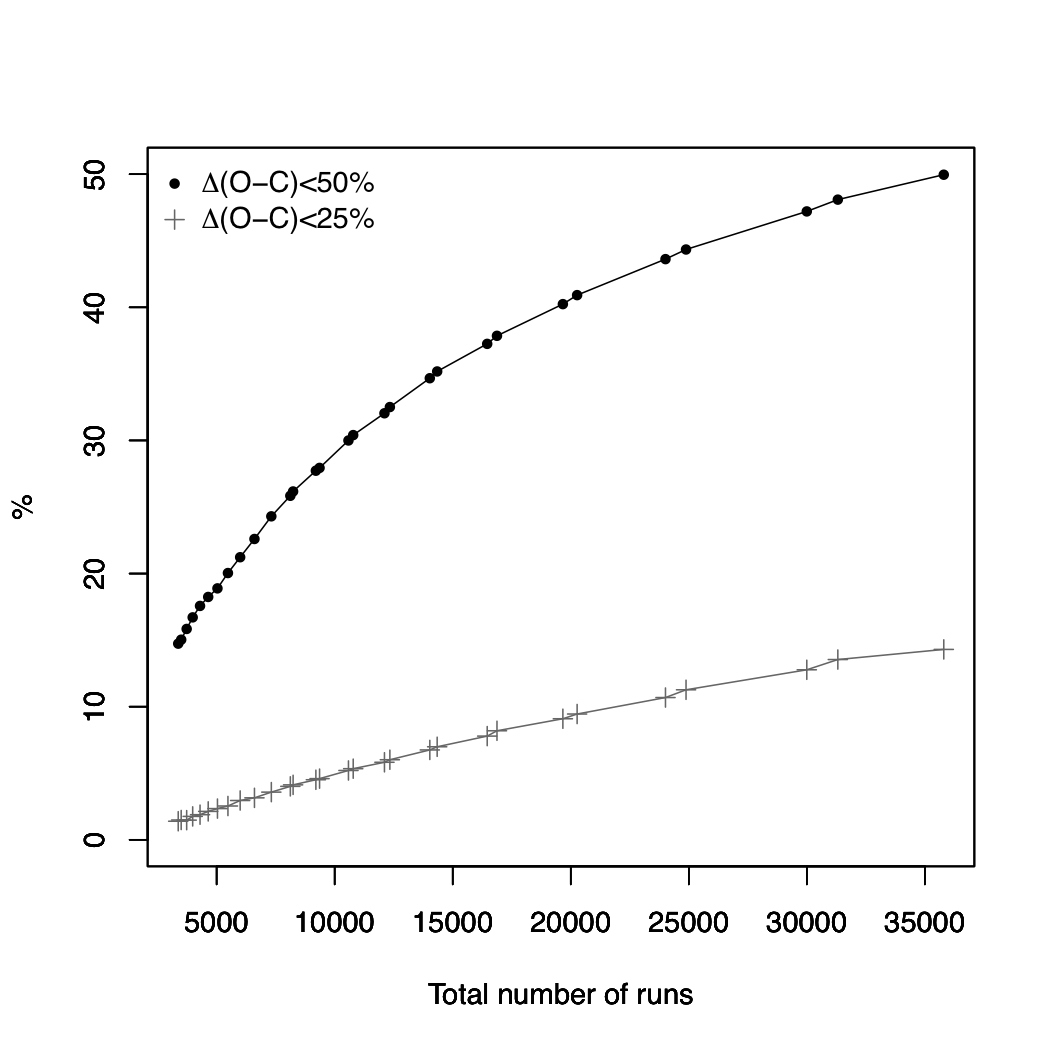}
\caption{Evolution of the number of selected ephemerides using the criteria of 50\% and of 25\% for  $\Delta (O-C)_{max}$ respectively.}
\label{pop5025}
\end{figure}

\begin{figure}
\includegraphics[width=8cm]{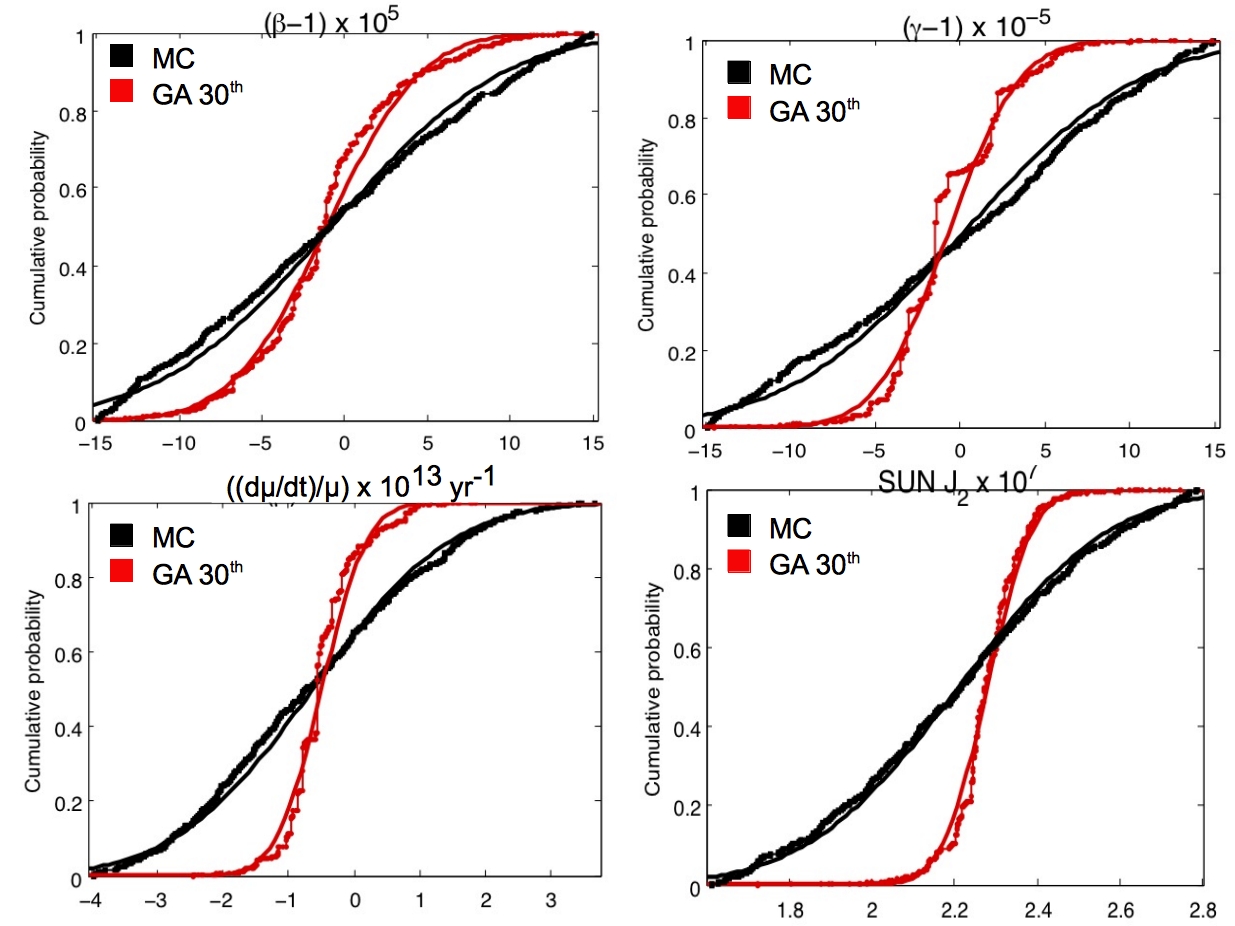}
\caption{Cumulative histograms of (PPN $\beta$, PPN $\gamma$, $J_{2}^\odot$, $\dot{\mu}/\mu$) for the generation 0 of ephemerides selected with the  $\Delta (O-C)_{max} < 50\% $ (noted MC and colored in black) and for the 30th generation (notes  GA  30$^{th}$ in red) also selected with $\Delta (O-C)_{max} < 50\% $. The full lines correspond to the cumulative histograms of the normal distributions fitted on the distributions of the first generation and the 30th generation.}
\label{qqplot}
\end{figure}

\begin{figure}
\includegraphics[width=9cm]{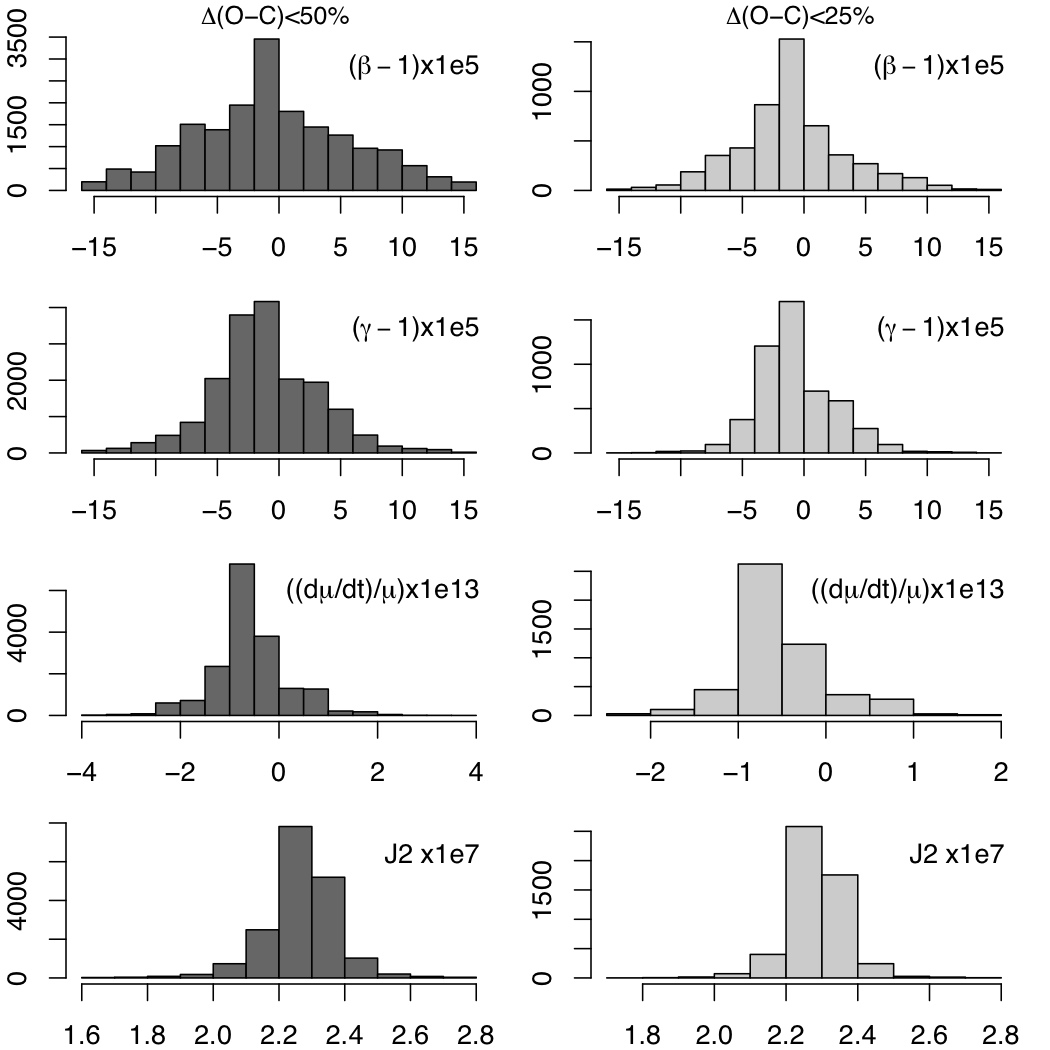}
\caption{Histograms of the distribution of the PPN $\beta$, PPN $\gamma$, $J_{2}^\odot$, and $\dot{\mu}/\mu$ parameters corresponding to the selected ephemerides using the criteria of 50\% and the 25\% differences of postfit residuals respectively, after 30 generations.}
\label{hist5025}
\end{figure}

\begin{figure}
\includegraphics[width=9cm]{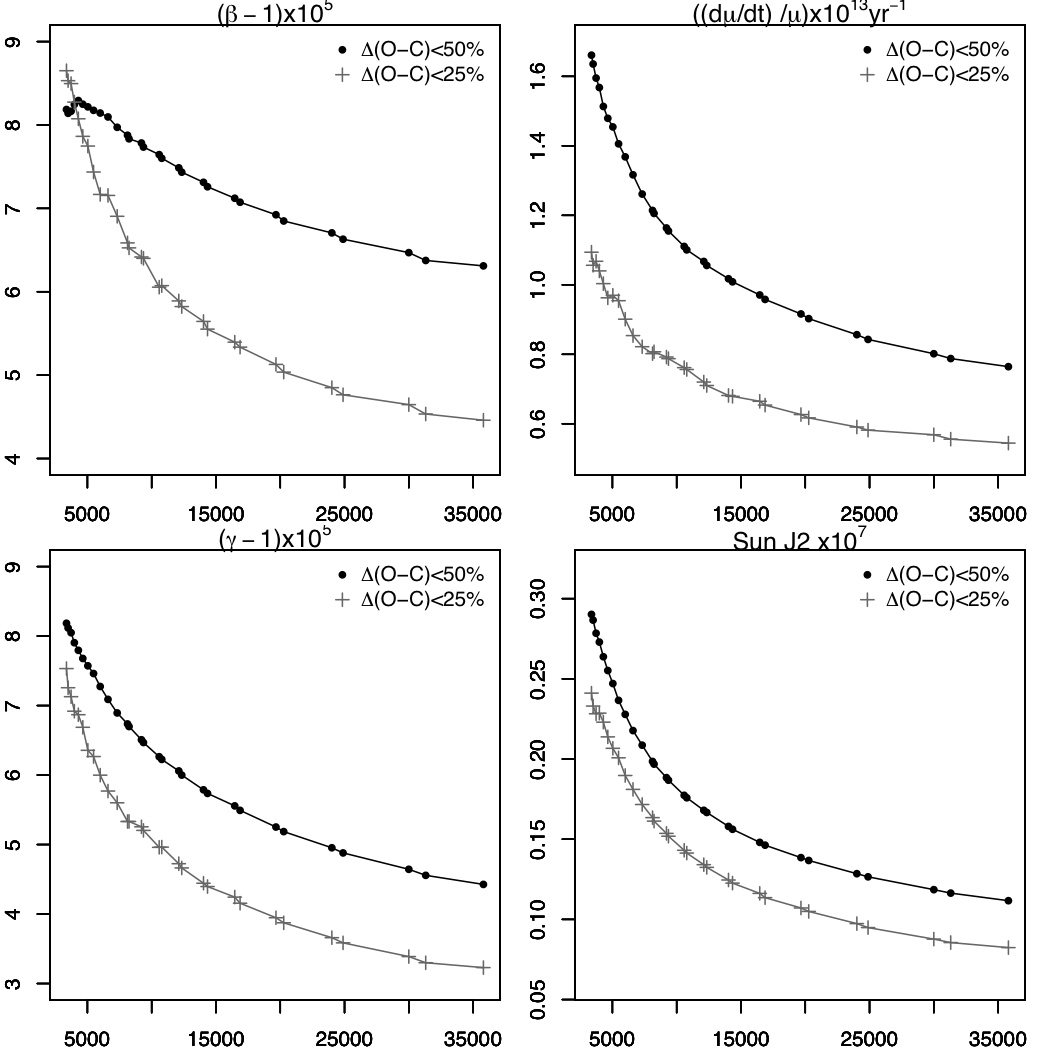}
\caption{Evolution with the total number runs of the 1-$\sigma$ of the gaussian distribution of the PPN $\beta$, PPN $\gamma$, $J_{2}^\odot$, and $\dot{\mu}/\mu$ parameters corresponding to the selected ephemerides using $\Delta (O-C)_{max} < 50\% $ and $\Delta (O-C)_{max} < 25\% $ respectively.}
\label{sig5025}
\end{figure}

On Figure \ref{pop5025} are given the percentages of runs with $\Delta (O-C)_{max} < 50\% $ and $\Delta (O-C)_{max} < 25\% $. 
The rates of selected ephemerides increase with the number of runs reaching 50 \% of the total runs for $\Delta (O-C)_{max} < 50\% $ and 14 \% for $\Delta (O-C)_{max} < 25\% $. 
The Gaussianity of the ($\beta$, $\gamma$, $J_{2}^\odot$, $\dot{\mu}/\mu$) samples corresponding to the selected ephemerides is also improved as one can see on the cumulative histograms of Figure \ref{qqplot} and on the histograms of Figure \ref{hist5025}.  
It is then possible to estimate mean values and $1-\sigma$ standard deviations of the sample of PPN $\beta$, $\gamma$, $J_{2}^\odot$ and $\dot{\mu}/\mu$ corresponding to the selected ephemerides. 
As one can see on Figure \ref{sig5025}, the dispersion of the selected samples of ($\beta$, $\gamma$, $J_{2}^\odot$, $\dot{\mu}/\mu$) decreases with the number of solutions and the percentages of selected ephemerides. 
The reduction of the dispersion value is generally of about a factor 2 from the generation 0 to the final generation. 

The mean and $\sigma$ obtained from the description of the distribution of the (PPN $\beta$, PPN $\gamma$, $J_{2}^\odot$, $\dot{\mu}/\mu$) samples corresponding to the selected ephemerides give the intervals of acceptable violation of general relativity on the basis of postfit residual differences to INPOP13c. 
 Table \ref{tab1} gathers these estimations labelled MC+ GA  50\% for the samples of parameters inducing $\Delta (O-C)_{max} < 50\% $ and MC+ GA  25\% for samples inducing $\Delta (O-C)_{max} < 25\% $, together with values found in the literature. 
 The obtained intervals are very competitive even if the intervals for $\beta$ and $\gamma$ are larger than those obtained with INPOP13a. This was expected as variations of $J_{2}^\odot$ and $\dot{\mu}/\mu$ were included in these results but not in INPOP13a.
 However other criteria can be used for selecting the ephemerides and for defining new thresholds.

\begin{table*}

\begin{center}
\caption{Results compared to values found in the literature.}
\label{tab1}
\begin{tabular}{l l c c c c c}
\hline
Method & Selected & PPN $\beta -1$ & PPN $\gamma -1$ & $\dot{\mu/\mu}$ & $\dot{G}$/G & J$_{2}^{\odot}$ \\
& runs & $\times$ 10$^{-5}$ & $\times$ 10$^{-5}$ & $\times$ 10$^{13}$ yr$^{-1}$ & $\times$ 10$^{13}$ yr$^{-1}$ & $\times$ 10$^{7}$ \\
\hline
\hline
{\bf{This paper}} & & & & & \\
MC & 3694 &-0.8 $\pm$ 8.2 & 0.2 $\pm$ 8.2 & -0.63 $\pm$ 1.66 & 0.04 $\pm$ -0.08 & 1.81 $\pm$ 0.29 \\
{MC +  GA  50 \%} & 17888 &{-0.49 $\pm$ 6.31} &{-1.19 $\pm$ 4.43} & -0.56 $\pm$ 0.76 &{-0.01 $\pm$ 0.91} &{2.26 $\pm$ 0.11} \\
{MC +  GA  25 \%} & 5120 &{-1.06 $\pm$ 4.46} &{-0.75 $\pm$ 3.23} & -0.51 $\pm$ 0.54 &{0.04 $\pm$ 0.69} &{2.28 $\pm$ 0.08} \\
\hline
MC +  GA  H$_{\textrm{RS}}$ & 8537 &{-0.01 $\pm$ 7.10} &{-1.67 $\pm$ 5.25} & -0.37 $\pm$ 0.76 &{ 0.18 $\pm$ 0.91} &{2.22 $\pm$ 0.14} \\
\hline
$<$ MC +  GA  (50 \% + RS) $>$ & & -0.25 $\pm$ 6.70& -1.48 $\pm$ 4.82 &-0.46$\pm$ 0.76 & 0.09 $\pm$ 0.91 & 2.24$\pm$ 0.125\\
\hline
\hline
{\bf{Planetary ephemerides}} & & & & & \\
DE (\cite{2011Icar..211..401K}) &&{4 $\pm$ 24} &fixed to (2.1 $\pm$ 2.3) & 0.0 & 0.0 & {fixed to 1.8} \\
&&fixed &{18 $\pm$ 26} &0.0 & 0.0 & {fixed to 1.8} \\
&& 0.0 &0.0 & 0.1 $\pm$ 1.6 & {0.65 $\pm$ 1.75}$^{*}$ & {fixed to 1.8} \\
DE (\cite{2014IPNPR.196C...1F}) & &0.0 & 0.0 & 0.0 & 0.0 & 2.1 $\pm$ 0.70 \\
EMP (\cite{2013MNRAS.432.3431P}) &&{-2 $\pm$ 3} &{4 $\pm$ 6} & & &{2.0 $\pm$ 0.2} \\
& & && -0.63 $\pm$ 0.43 &{-0.08 $\pm$ 0.58}$^{*}$ & \\
{INPOP13a (\cite{2014A&A...561A.115V})} & &{0.2 $\pm$ 2.5} &{-0.3 $\pm$ 2.5} & 0.0 & {0.0} &{ 2.40 $\pm$ 0.20}  \\
INPOP10a (\cite{Fienga2011}) &&  -4.1 $\pm$ 7.8 & -6.2 $\pm$ 8.1& 0.0 & 0.0 & 2.40 $\pm$ 0.25\\
INPOP08 (\cite{2009AA...507.1675F}) && 7.5 $\pm$ 12.5 & 0.0 & 0.0 & 0.0& 1.82 $\pm$ 0.47 \\
\hline
{\bf{LLR }} & & & & & \\
\cite{2009IJMPD..18.1129W} && 12 $\pm$ 11 & fixed to (2.1 $\pm$ 2.3) & & 0.0 & \\
\cite{2009IAU...261.0801W} && 0.0 & 0.0 & & $\pm$ 3 & \\
\cite{2010AA...522L...5H} & & 0.0 & 0.0 & & {-0.7 $\pm$ 3.8} &{} \\
& & 3 $\pm$ 13 & fixed to (2.1 $\pm$ 2.3)& & 0.0 & \\
\hline
{\bf{Other technics }} & & & & & \\
Cassini (\cite{2003Natur.425..374B}) && 0.0 &{2.1 $\pm$ 2.3} & 0.0 &0.0 & NC \\
VLBI (\cite{2009AA...499..331L}) && 0.0 & -8 $\pm$ 12 & 0.0 &0.0 & fixed \\
Planck + Brans-Dicke (\cite{2013PhRvD..88h4053L}) && & &&{-1.315$\pm$ 2.375} & \\
Binary pulsar (\cite{kaspi94}) && & &&{40$\pm$ 50} &\\ 
Heliosismology  (\cite{guenther98}) && & &&{0$\pm$ 16} &\\ 
Big Bang nucleosynthesis	(\cite{bambi05}) && & && 0 $\pm$ 4 &\\
\hline
\end{tabular}
\end{center}
{\it{$^{*}$ This value is not the published one but is calculated in using the value of $\dot{M}/M$ given in section \ref{gdot}}}

\end{table*} 
 
\section{The resampling criteria}
\label{GAdisc}
\subsection{Definition}
\begin{figure}
\includegraphics[width=8cm]{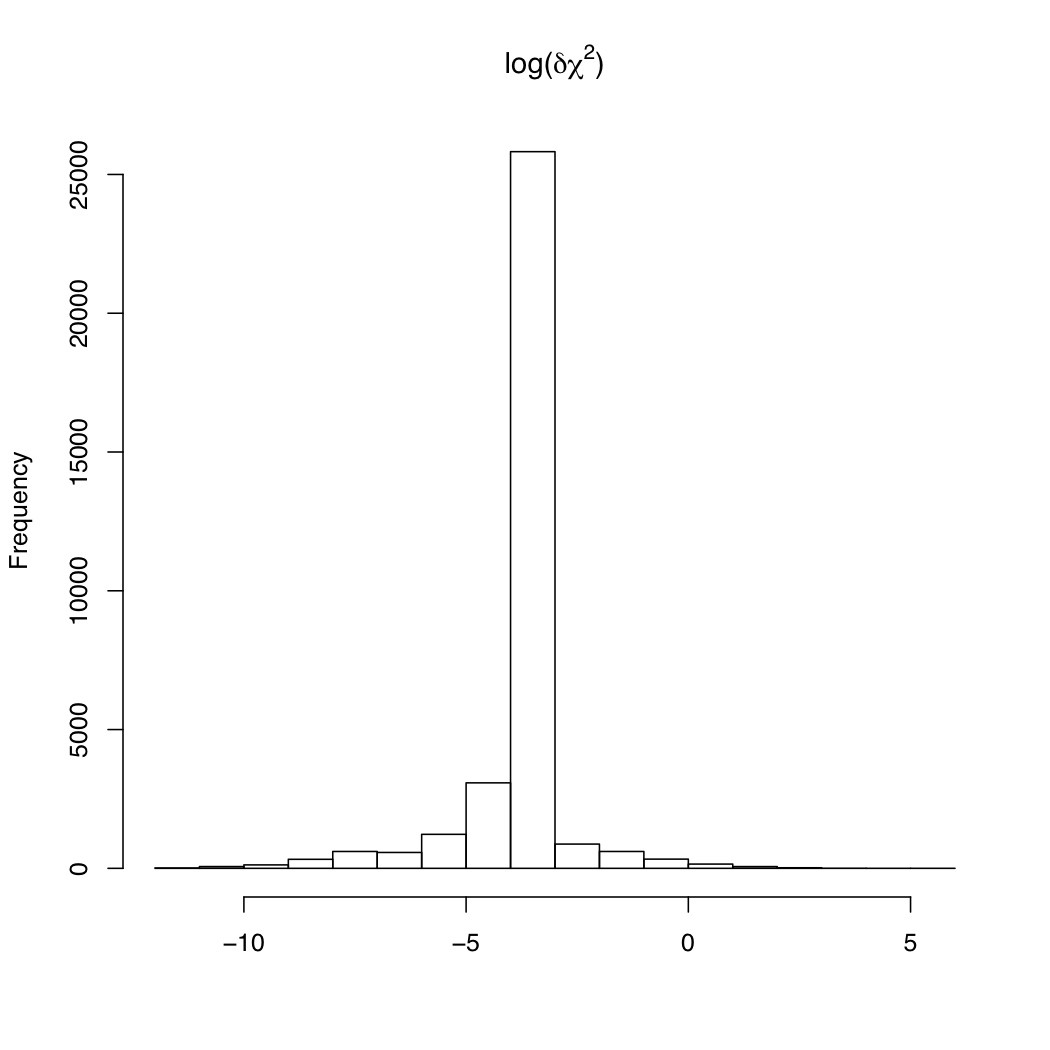}
\caption{Histogram of the log of the differences $\delta \chi^{2}$  between the $\chi^{2}$ obtained for INPOP13c and the $\chi^{2}$ computed during the construction of the random (PPN $\beta$, PPN $\gamma$, $J_{2}^\odot$, $\dot{\mu}/\mu$) ephemerides.}
\label{histochi2}
\end{figure}

The differences between two least squares adjustments can be characterized by the $\chi^{2}$ values obtained for each fit.
The $\chi^{2}$ is usually given by
$$
{\rm{(N-M)}} \, \chi^{2} = \sum_{i=1}^{N} \frac{(O - C)_{i}^{2}}{\sigma_{i}^{2}}
$$
where $N$ is the number of data used for the adjustment, $M$ the number of fitted parameters, $\sigma_{i}$ the estimated variance of the observation $i$ and $(O - C)_{i}$ is the postfit residual of the data $i$.
In the theory of least squares, the minimization of $\chi^{2}$ is the main criteria for the convergency of the adjustment. Considering the difference between two ephemerides by their $\chi^{2}$ differences, one characterizes how well the two dynamical modelings represent the weighted observations. 
We have then compared the $\chi^{2}$ obtained with a random (PPN $\beta$, PPN $\gamma$, $J_{2}^\odot$, $\dot{\mu}/\mu$) ephemerides with the $\chi^{2}$ of INPOP13c. We note $\delta \chi^{2}$ such difference with:
$$
\delta \chi^{2} = \chi^{2}_{(\beta, \gamma, J_{2}^\odot, \dot{G}/G)} - \chi^{2}_{\textrm{IN}}
$$
The histogram of the obtained differences noted $\delta \chi^{2}$ is plotted on Figure \ref{histochi2}. As one can see, the differences can reach several orders of magnitude but can also be very small. 
In order to make a selection of which differences are negligible compared to the INPOP13c uncertainties, we use a method based on the resampling of the observational data sets as the observational uncertainties are the main source of error for the planetary ephemerides (section \ref{jack}). 
 
\subsection{Resampling method}
\label{jack}
One way to evaluate the planetary ephemeris uncertainties in terms of $\chi^{2}$ is to consider the differences of the $\chi^{2}$ induced by the variances and bias of the observational data sets used for the fit. 
An interesting tool is then to use a resampling technique. 
By removing randomly a part of the observational sample used for the fit (\cite{Busing97}), one can estimate the robustness of the estimated parameters but also a realistic estimation of the postfit residuals and of the $\chi^{2}$(\cite{cook82}, \cite{Fay85}). 
The variations of the $\chi^{2}$ will indicate the sensibility of the fit to the observational sampling.
We operate 101 runs by removing randomly 10\% of the INPOP data sample. We then estimate the variation of the $\chi^{2}$ in comparison to the $\chi^{2}$ obtained for INPOP13c. We note:
$$
\Delta \chi^{2}_{j} = \chi^{2}_{j} - \chi^{2}_{\textrm{IN}}
$$
with $j=1,...101$.
 Such $\Delta \chi^{2}$  gives an estimation of the variation of the $\chi^{2}$ induced by the variance and bias in the data sampling.  We keep as a threshold  the maximum value of $\Delta \chi^{2}$ obtained after 101 tests of random modification of the data sample and labelled $\Delta \chi^{2}_{max}$. 
It gives the limit for the $\delta \chi^{2}$ differences: a planetary ephemerides built with a random selection of (PPN $\beta$, PPN $\gamma$, $J_{2}^\odot$, $\dot{\mu}/\mu$) is close to INPOP13c if the obtained $\chi^{2}$ does not differ from the INPOP13c $\chi^{2}$ by more than the maximum $\chi^{2}$ differences induced by the data sample bias and variances. In the following, this test will be called the resampling criteria or noted H$_{\textrm{RS}}$.


\begin{figure}
\includegraphics[width=8cm]{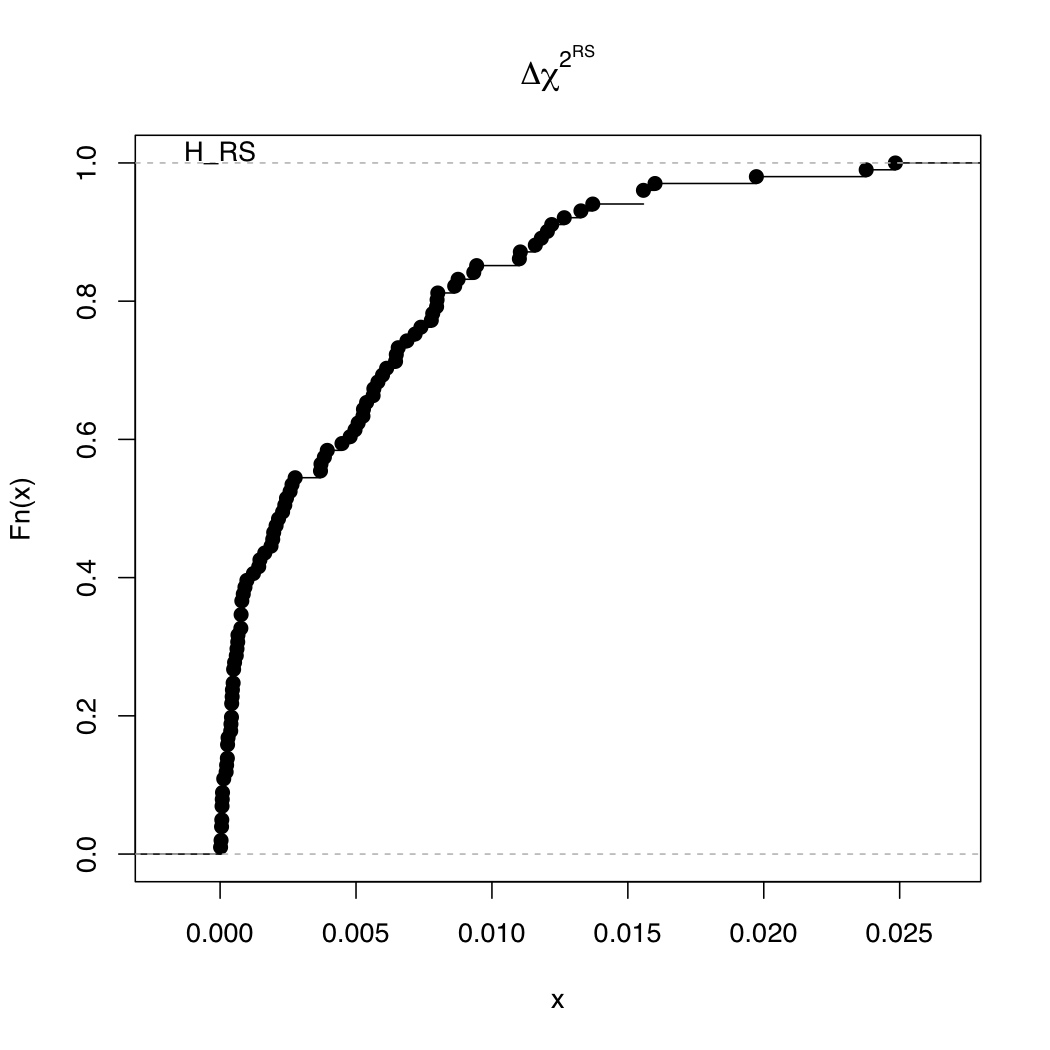}
\caption{Cumulative histogram of $\Delta \chi^{2}$ obtained for the 101 resampled runs.}
\label{ecdf}
\end{figure}

\begin{figure}
\includegraphics[width=8cm]{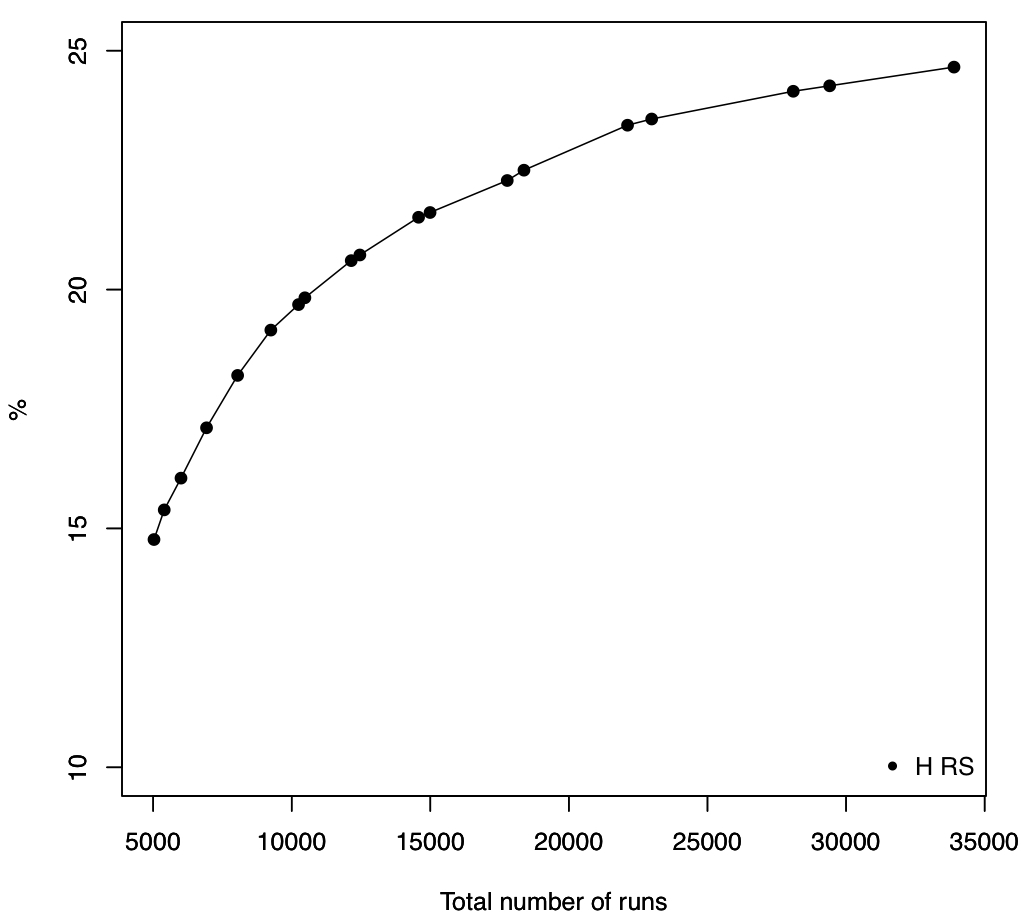}
\caption{Evolution of the number of selected ephemerides based on the resampling criteria described in the text.}
\label{num4}
\end{figure}

\begin{figure}
\begin{center}
\includegraphics[width=8cm]{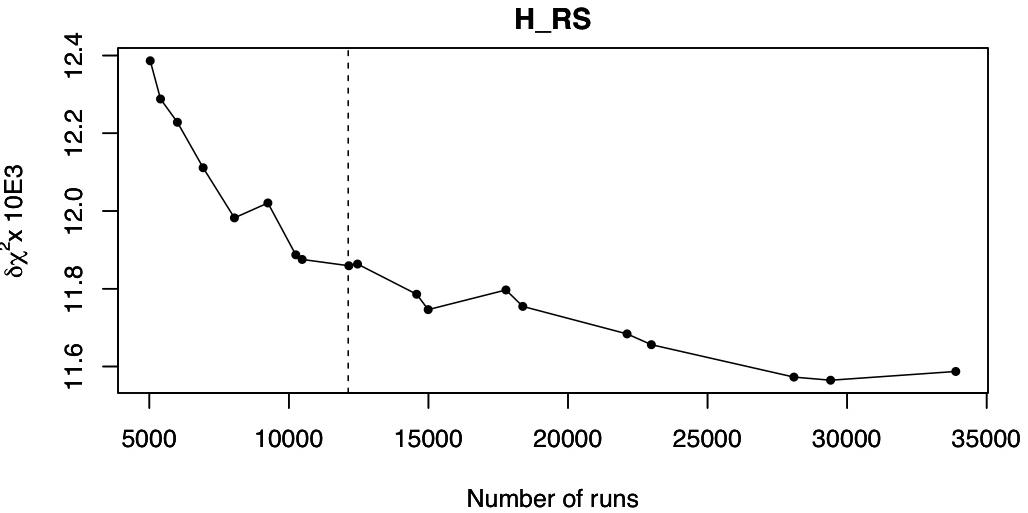}
\end{center}
\caption{Evolution of the average values of $\delta \chi^{2}$ (y-axis) regarding the total number of ephemerides (x-axis) selected with the resampling criteria. }
\label{convchi2}
\end{figure}

\begin{figure}
\includegraphics[width=4.5cm]{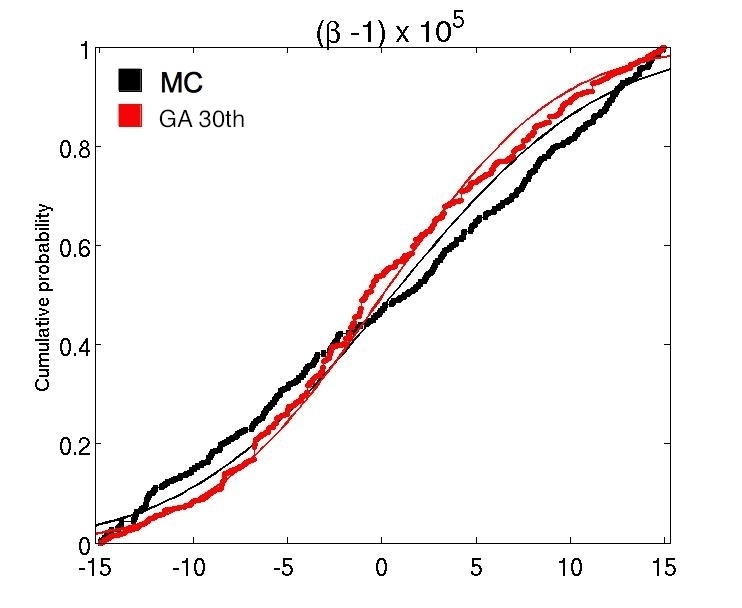}\includegraphics[width=4.5cm]{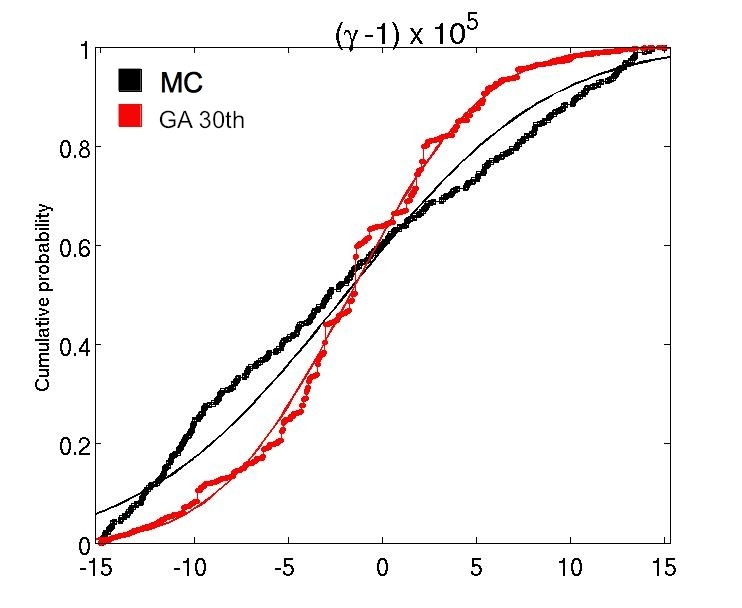}\\
\includegraphics[width=4.5cm]{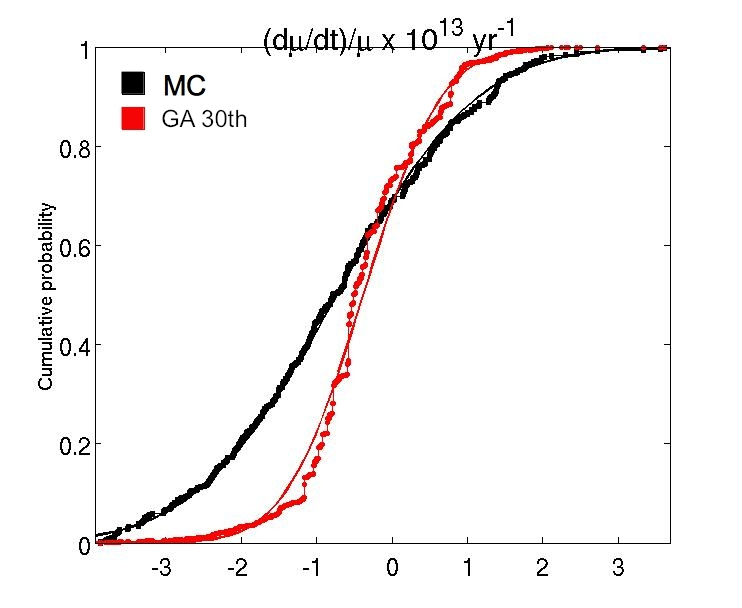}\includegraphics[width=4.5cm]{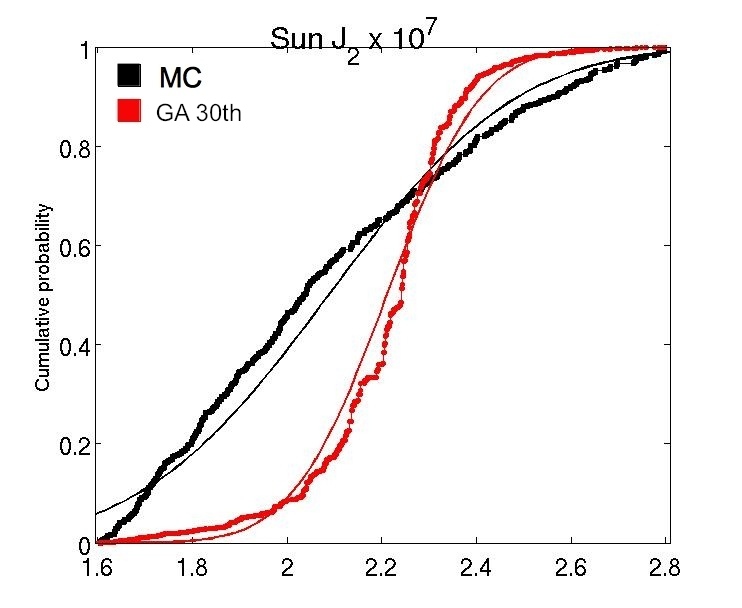}
\caption{Cumulative histogram of (PPN $\beta$, PPN $\gamma$, $J_{2}^\odot$, $\dot{\mu}/\mu$) for the generation 0 of ephemerides selected with the  resampling criteria (noted MC and colored in black) and for the final generation noted GA 30$^{th}$ and colored in red also selected with the resampling criteria. The full lines are the corresponding cumulative histograms for the normal distribution fitted on the distributions of the first generation and the 30th generation.}
\label{cumul}
\end{figure}

\begin{figure}
\includegraphics[width=9cm]{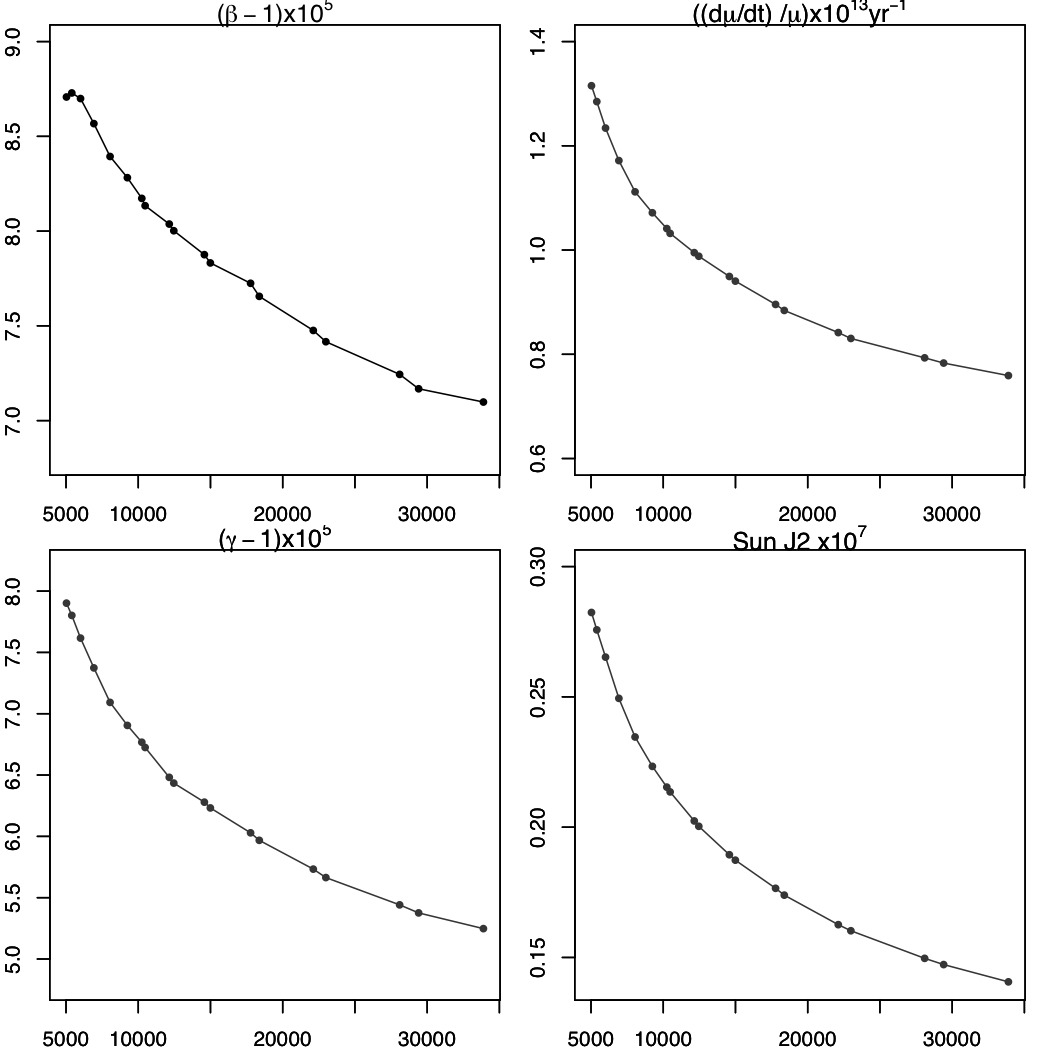}
\caption{Evolution with the number of selected runs of the 1-$\sigma$ of the gaussian distribution of the PPN $\beta$, PPN $\gamma$, $J_{2}^\odot$, and $\dot{\mu}/\mu$ parameters corresponding to the ephemerides selected with the resampling criteria.}
\label{sigH4}
\end{figure}

\begin{figure}
\includegraphics[width=9cm]{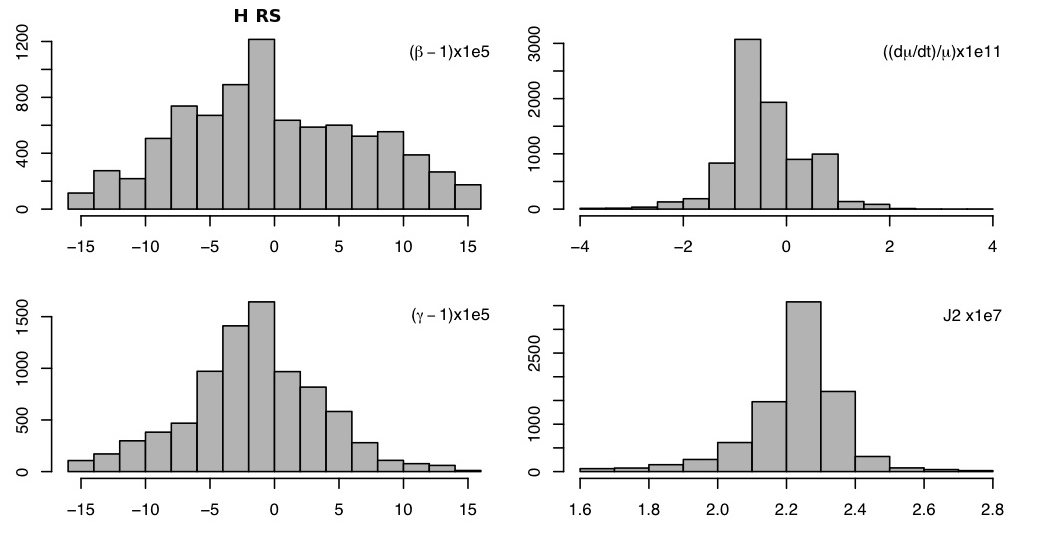}
\caption{Histograms of the distribution of the PPN $\beta$, PPN $\gamma$, $J_{2}^\odot$, and $\dot{\mu}/\mu$ parameters corresponding to the ephemerides selected with the resampling criteria for the full sample of simulated ephemerides.}
\label{H4}
\end{figure}


%
\subsection{Results}
\label{GAresults2}

On figure \ref{num4}, are plotted the number of selected random (PPN $\beta$, PPN $\gamma$, $J_{2}^\odot$, $\dot{\mu}/\mu$) ephemerides based on the resampling criteria. As one can see, the number of selected ephemerides increases with the number of runs. 
The convergence of the algorithm is illustrated on Figure \ref{convchi2} where the average values of the $\delta \chi^{2}$ for each generation are plotted regarding the cumulated number of runs. The stabilization is obtained before the 35800 runs (30th generation).

As one can see on figure \ref{cumul}, the Gaussianity of the (PPN $\beta$, PPN $\gamma$, $J_{2}^\odot$, $\dot{\mu}/\mu$) parameters corresponding to the selected ephemerides improves with the runs. 
Furthermore, on figure \ref{sigH4}, it also appears that the uncertainties deduced from these gaussian distributions of (PPN $\beta$, PPN $\gamma$, $J_{2}^\odot$, $\dot{\mu}/\mu$) also decrease with the number of runs.
For illustration we give on figure \ref{H4} the histograms of the (PPN $\beta$, PPN $\gamma$, $J_{2}^\odot$, $\dot{\mu}/\mu$) parameters deduced for the full analysis of the 35800 runs (30$^{th}$ generation). 
The mean values and 1-$\sigma$ standard deviations of (PPN $\beta$, PPN $\gamma$, $J_{2}^\odot$, $\dot{\mu}/\mu$) parameters extracted from these histograms are given on Table \ref{tab1}.


In general, in comparison to the previous criteria of ephemeris selection based on the maximum differences in the postfit residuals (see section \ref{3.2}), the criteria based on the estimation of the variations of the $\chi^{2}$ are more selective. 
But in the same time, the gaussian distributions of the (PPN $\beta$, PPN $\gamma$, $J_{2}^\odot$, $\dot{\mu}/\mu$) parameters are slightly larger leading to less restrictive intervals of possible violation of general relativity.

\section{Discussions}

On Table \ref{tab1} are gathered the results obtained with this work as well as very diverse estimations found in the literature.  
If one wants to exhibit one single set of values of acceptable intervals for the four parameters randomly modified in this work, one can consider the mean values of the most numerous selection presented in Table \ref{tab1}, gathering values of (PPN $\beta$, PPN $\gamma$, $J_{2}^\odot$, $\dot{\mu}/\mu$) inducing ephemerides with  $\Delta (O-C)_{max} < 50\%$ and ephemerides selected with the resampling criteria. 
By considering the mean of these two selections, we then obtain the values labeled  MC +  GA  (50 \% + RS) in Table \ref{tab1}.

As noticed in \cite{2014A&A...561A.115V} the interval of possible violations for the PPN parameters $\beta$ and $\gamma$  with no time variation of the Newtonian gravitational constant G and in fixing the value of the Sun flattening is as accurate as the reference values obtained with the Cassini experiment (\cite{2003Natur.425..374B}).

However by adding the variations of $\dot{\mu}/\mu$ and $J_{2}^\odot$, we have enlarge the possible interval of violations for the four parameters as given in line MC +  GA  (50 \% + RS) of Table \ref{tab1}. 
This result is consistent with the fact that these parameters are linked through their influences on the orbital elements of the planets as described by Equation \ref{perih}. 
In opposition to \cite{2013MNRAS.432.3431P}, $\beta$ intervals are always larger than $\gamma$ intervals. 
This can be explained again by the Equation \ref{perih}. 
The determination of PPN $\beta$ can only be done with this equation where $\dot{\mu}/\mu$ and $J_{2}^\odot$ influences can mix up with those of  $\beta$ and $\gamma$.
In the meantime, PPN $\gamma$ can be constrained by the Shapiro delay where only $\dot{\mu}/\mu$ plays a role with $\gamma$. 
The $\beta$ intervals deduced from LLR analysis are still larger than those obtained in this work. 
The LLR determinations are based only on the modeling of the Moon orbit about the Earth. 
As explained in introduction, the planetary ephemerides provide better determinations of PPN parameters thanks to the eight equations of motion considered in the computations. 
The acceptable interval of $\dot{\mu}/\mu$ deduced from this work is surprisingly a factor 2 more narrow than the one deduced from LLR analysis. 
This can again be explained by the disentangling operated between the gravitational signatures of PPN $\beta$, PPN $\gamma$, $J_{2}^\odot$, $\dot{\mu}/\mu$ by considering the eight planetary equations of motion simultaneously.
Comparisons to \cite{2013MNRAS.432.3431P} seem to be difficult as no clear evidence of a simultaneous fit of the four considered parameters is given in their paper. 
In the other hand, as \cite{2013MNRAS.432.3431P} values are directly extracted from a least square procedure, we can expect to have small uncertainties from the least square determinations in comparison to our statistic intervals of violation including all correlations.
Values extracted from \cite{2011Icar..211..401K} were obtained with planetary ephemerides built before the use of Messenger tracking data in their construction. The subsequent improvement of the Mercury orbit is then not included in \cite{2011Icar..211..401K} results but in \cite{2014IPNPR.196C...1F} and its value of $J_{2}^\odot$. 
Future JPL estimations of PPN parameters and $\dot{\mu}/\mu$ should also benefit from the improved Mercury orbit.

Finally, values obtained by astrophysical technics such as heliosismology or pulsar timing analysis are less restrictive than those obtained in the solar system.
However, considering all the given figures of Table \ref{tab1} one should conclude that no deviation to general relativity is noticeable for the four parameters modified simultaneously. 

\section{Conclusions}

In this work we have estimated new limits of possible violations of general relativity with the PPN parameters $\beta$, $\gamma$ in considering in the same time time variations of the Gravitational constant G and various values of the sun flattening. We used Monte Carlo simulations and genetic algorithm procedures for producing more than 35000 planetary ephemerides fitted to observations and compared to INPOP13c. 
Different criteria have been used for characterizing the closest ephemerides from INPOP13c and so, for identifying the most acceptable intervals of parameters inducing the smallest modifications to the planetary dynamics.

New tests will be implemented such as the addition of supplementary terms in the equation of motions of the planets as proposed by alternative theories (\cite{2011MNRAS.412.2530B}, \cite{2014PhRvD..89j2002H}, \cite{2011mmgr.book..491J}). Tests of the equivalence principal can also be proposed for Monte Carlo simulations and genetic algorithm procedures. In the case of the planetary orbits, one would have to consider one ratio of gravitational and inertial masses for each planet which would multiply the number of runs by an important scale.

\section*{Acknowledgments}
This work benefited from HPC resources of MesoPSL financed by the Region Ile de France and the project Equip@Meso (reference ANR-10-EQPX-29-01) of the programme Investissements d'Avenir supervised by the Agence Nationale pour la Recherche. The authors thank also Professor Damour for his fruitfull discussions.

\bibliographystyle{aa}
\bibliography{biblio_hdr}

\begin{appendix}
\section{INPOP13c data sample and postfit residuals}

\begin{table*}
\caption{Statistics (means and 1-$\sigma$ standard deviations) of the residuals obtained after the INPOP13c fit for common data sample between INPOP13c and INPOP10e. }
\begin{center}
\begin{tabular}{l c l c | c c | c c }
\hline
Type of data & & Nbr & Time Interval & \multicolumn{2}{c}{INPOP10e}& \multicolumn{2}{c}{INPOP13c} \\
\hline
Mercury & range [m]& 462 & 1971.29 - 1997.60 & -45.3 & 872.5 &  -101.5 &    861.5  \\
Mercury  Mariner & range [m]& 2 & 1974.24 - 1976.22 & -31.8 &  109.2 & -196.4 &     19.6   \\
Mercury  flybys  Mess & ra [mas]& 3 & 2008.03 - 2009.74 & 0.7  & 1.5 & 0.9 &      1.3   \\
Mercury  flybys  Mess & de [mas]& 3 & 2008.03 - 2009.74 & 2.4  & 2.5 & 2.5 &       2.4  \\
Mercury  flybys  Mess & range [m]& 3 & 2008.03 - 2009.74 & -5.1 & 5.8  & 3.2 &       7.7   \\
\hline
Venus & VLBI [mas]& 46 & 1990.70 - 2010.86 & 1.6 & 2.6  & 1.6 &      2.6    \\
Venus & range [m]& 489 & 1965.96 - 1990.07 & 500.2 & 2234.9  & 504.6 &   2237.6   \\
Venus  Vex & range [m]& 22145 & 2006.32 - 2009.78 & -0.0 & 4.1  &  1.0 &      5.1  \\
\hline
Mars & VLBI [mas]& 96 & 1989.13 - 2007.97 & -0.0 & 0.4  &  0.0 &      0.4 \\
Mars  Mex & range [m]& 13842 & 2005.17 - 2009.78 & 0.4 & 3.2 &  -0.5 &      1.8  \\
Mars  MGS & range [m]& 13091 & 1999.31 - 2006.83 & -0.3 & 3.8  & 0.4 &      3.8   \\
Mars  Ody & range [m]& 5664 & 2006.95 - 2010.00 & 0.3 & 4.1  &   1.5 &      2.3  \\
Mars  Path & range [m]& 90 & 1997.51 - 1997.73 & -6.3 & 13.7  & 19.3 &     14.1   \\
Mars  Vkg & range [m]& 1257 & 1976.55 - 1982.87 & -1.4 & 39.7  & -1.5 &     41.2   \\
\hline
Jupiter & VLBI [mas]& 24 & 1996.54 - 1997.94 & -0.3 & 11.068  & -0.450 &     11.069    \\
Jupiter & ra [mas]& 6532 & 1914.54 - 2008.49 & -39.0 & 297.0 &  -39.0 &      297.0   \\
Jupiter & de [mas]& 6394 & 1914.54 - 2008.49 & -48.0 & 301.0  & -48.0 & 301.0   \\
Jupiter  flybys & ra [mas]& 5 & 1974.92 - 2001.00 & 2.4 & 3.2  & 2.5 &      3.0   \\
Jupiter  flybys & de [mas]& 5 & 1974.92 - 2001.00 & -10.8 & 11.5 & -10.8 &      11.4  \\
Jupiter  flybys & range [m]& 5 & 1974.92 - 2001.00 & -907.0 & 1646.2 &  -986.0 &    1775.6  \\
\hline
Saturne & ra [mas]& 7971 & 1913.87 - 2008.34 & -6.0 & 293.0 &  -6.0 & 293.0  \\
Saturne & de [mas]& 7945 & 1913.87 - 2008.34 & -12.0 & 266.0 & -12.0 & 266.0  \\
Saturne  VLBI  Cass & ra [mas]& 10 & 2004.69 - 2009.31 & 0.215 & 0.637 &  0.113 &      0.630   \\
Saturne  VLBI  Cass & de [mas]& 10 & 2004.69 - 2009.31 & 0.280 & 0.331 &   -0.115 &      0.331 \\
Saturne  Cassini & ra [mas]& 31 & 2004.50 - 2007.00 & 0.790 & 3.879 &  0.663 &      3.883  \\
Saturne  Cassini & de [mas]& 31 & 2004.50 - 2007.00 & 6.472 & 7.258 &   5.906 &       7.284 \\
Saturne  Cassini & range [m]& 31 & 2004.50 - 2007.00 & -0.013 & 18.844 &   0.082 &      23.763   \\
\hline

Uranus & ra [mas]& 13016 & 1914.52 - 2011.74 & 7.0 & 205.0 & 7.0 & 205.0  \\
Uranus & de [mas]& 13008 & 1914.52 - 2011.74 & -6.0 & 234.0 & -6.0 & 234.0   \\
Uranus  flybys & ra [mas]& 1 & 1986.07 - 1986.07 & -21.0 & 0.000 & -21.0 & 0.000  \\
Uranus  flybys & de [mas]& 1 & 1986.07 - 1986.07 & -28.0 & 0.000 & -28.0 & 0.000    \\
Uranus  flybys & range [m]& 1 & 1986.07 - 1986.07 & 19.7 & 0.000 &20.8 & 0.000  \\
\hline
Neptune & ra [mas]& 5395 & 1913.99 - 2007.88 & 0.0 & 258.0 & 3.0 & 258.0 \\
Neptune & de [mas]& 5375 & 1913.99 - 2007.88 & -0.0 & 299.0 & -2.0 & 299.0  \\
Neptune  flybys & ra [mas]& 1 & 1989.65 - 1989.65 & -12.0 & 0.000 & -11.0& 0.000    \\
Neptune  flybys & de [mas]& 1 & 1989.65 - 1989.65 & -5.0 & 0.000 & -5.0 & 0.000    \\
Neptune  flybys & range [m]& 1 & 1989.65 - 1989.65 & 69.6 & 0.000 & 51.5& 0.000     \\
\hline
Pluto & ra [mas]& 2458 & 1914.06 - 2008.49 & 34.0 & 654.0 & 20.0 & 574.0     \\
Pluto & de [mas]& 2462 & 1914.06 - 2008.49 & 7.0 & 539.0 & 1.0 & 525.0     \\
Pluto  Occ & ra [mas]& 13 & 2005.44 - 2009.64 & 3.0 & 47.0 & -100.0 &  44.0     \\
Pluto  Occ & de [mas]& 13 & 2005.44 - 2009.64 & -6.0 & 18.0 &  0.0 &  27.0  \\
Pluto  HST & ra [mas]& 5 & 1998.19 - 1998.20 & -33.0 & 43.0 & -18.0 & 44.0     \\
Pluto  HST & de [mas]& 5 & 1998.19 - 1998.20 & 28.0 & 48.0 & -26.0 & 48.0   \\
\hline
\hline
\end{tabular}
\end{center}
\label{omctab1}
\end{table*}

\begin{table*}
\caption{Statistics of INPOP13c  postfit residuals for new samples included in the fit. For comparison, means and standard deviations of residuals obtained with INPOP10e on these prolongated intervals are also given.}
\begin{center}
\begin{tabular}{l c l c | c c | c c }
\hline
Type of data & & Nbr & Time Interval & \multicolumn{2}{c}{INPOP10e}& \multicolumn{2}{c}{INPOP13c} \\
\hline
Mercure Messenger & range [m]& 371 &  2011.39 - 2013.20 &  7.2   & 189.7   &  4.0 &     12.4\\
\hline
Venus Vex & range [m]& 2825 &  2009.78 - 2011.45 & 1.8 &  16.5 &    5.1 &     15.7 \\
\hline
Mars Mex & range [m]& 12268 &  2009.78 - 2013.00 &   3.6  &  23.3   & 1.2 &     5.6 \\
Mars Ody & range [m]& 3510 &  2010.00 - 2012.00 &    4.8   &  9.6    & 0.7&     1.8 \\
\hline
\end{tabular}
\end{center}
\label{omcsupp}
\end{table*}

\end{appendix}

\end{document}